\documentclass[letterpaper,english,aps,prl,floatfix,reprint,superscriptaddress,showpacs,twocolumn]{revtex4-1}
\usepackage[LGR,T1]{fontenc}
\usepackage[utf8]{inputenc}
\usepackage{bm}
\usepackage{amsmath}
\usepackage{amssymb}
\usepackage{stmaryrd}
\usepackage{graphicx}

\makeatletter

\pdfpageheight\paperheight
\pdfpagewidth\paperwidth

\DeclareRobustCommand{\greektext}{%
  \fontencoding{LGR}\selectfont\def\encodingdefault{LGR}}
\DeclareRobustCommand{\textgreek}[1]{\leavevmode{\greektext #1}}

\pdfoutput=1
\usepackage{mathrsfs}
\usepackage{bm}
\usepackage{xspace}
\usepackage{epstopdf}
\usepackage{longtable}
\usepackage{amsmath}    
\allowdisplaybreaks[4]  

\usepackage[colorlinks,linkcolor=blue,anchorcolor=blue,citecolor=blue]{hyperref}

\makeatother

\usepackage{babel}
\begin{document}
\title{Universal classes of disorder scattering for the in-plane anomalous
Hall effect}
\author{Guoao Yang}
\affiliation{School of Physics and Optoelectronics Engineering, Anhui University,
Hefei, Anhui Province 230601, P.R. China}
\affiliation{Anhui Provincial Key Laboratory of Low-Energy Quantum Materials and
Devices, High Magnetic Field Laboratory, HFIPS, Chinese Academy of
Sciences, Hefei, Anhui 230031, China}
\author{Tao Qin}
\email{taoqin@ahu.edu.cn}

\affiliation{School of Physics and Optoelectronics Engineering, Anhui University,
Hefei, Anhui Province 230601, P.R. China}
\author{Jianhui Zhou}
\email{jhzhou@hmfl.ac.cn}

\affiliation{Anhui Provincial Key Laboratory of Low-Energy Quantum Materials and
Devices, High Magnetic Field Laboratory, HFIPS, Chinese Academy of
Sciences, Hefei, Anhui 230031, China}
\date{\today}
\begin{abstract}
The in-plane anomalous Hall effect (IPAHE) with planar Hall current
and magnetization/magnetic fields in various quantum materials has
received increasing attention. Most of the current efforts are devoted
to the intrinsic part due to the Berry curvature of electronic bands,
however, how disorder scattering affects the extrinsic part (the skew
scattering and side jump) remains largely elusive. Here we theoretically
investigate the three universal classes of disorder scattering (scalar,
spin-conserving, and spin-flipping) for the IPAHE, based on the prototypical
two-dimensional massive Dirac fermion model with warping term under
generic Zeeman fields. We find that the different disorder scattering
results in a distinct dependence of the anomalous Hall conductivity
on disorder strength, and we recover previously known results within
some limits. Remarkably, the spin-flipping scattering could give rise
to nontrivial contributions featuring sinusoidal oscillations with
periods of \textgreek{π} and 2\textgreek{π} to the extrinsic part,
in contrast to the standard two-dimensional massive Dirac fermions.
Our work unveils the rich features of anomalous transport in planar
Hall geometry in the presence of disorder scattering and provides
some useful insights into the magnetotransport phenomena.
\end{abstract}
\maketitle

\section{Introduction}

In contrast to the conventional anomalous Hall effect (AHE) with perpendicular
Hall current and magnetization \citep{Nagaosa2010RMP}, the in-plane
anomalous Hall effect (IPAHE) in which the Hall current and the magnetization
are in the same plane has attracted considerable attention due to
the promising applications in low-energy electronics \citep{Malshukov1998prb,ZhangYP2011PRB,LiuX2013PRL,RenYF2016PRB,LiuZ2018PRL,You2019PRB,ZhangJL2019PRB,GeJ2020NSR,Zyuzin2020PRB,Cullen2021PRL,TanHX2021PRB,zhouJD2022nature,LiZY2022PRL,CaoJ2023PRL,Kurumaji2023PRR,Xue2023NSR,MiaoWQ2024PRB}.
It originates from the Berry curvature of electrons and is generally
attributed to the interplay between the magnetization and the spin-orbit
interaction in magnetic materials \citep{Xiao2010RMP}. Remarkably,
the IPAHE was experimentally observed in heterodimensional superlattice
$\mathrm{VS}-\mathrm{VS_{2}}$ nanoflakes induced by an in-plane magnetic
field \citep{zhouJD2022nature,CaoJ2023PRL}. This hasIt further inspireds
the search for new materials hosting IPAHE and its detection in novel
quantum materials \citep{CUI2024SB,WangH2024PRL,Xiao2025,pengWZ2024,Nakamura2024PRL,liu25}.

Disorder scattering significantly modifies a quasiparticle's lifetime
and phase, as well as its velocity matrix elements, which play a vital
role in the transport properties of electrons in solids \citep{LeePA1985RMP}.
It is known that disorder can significantly change the extrinsic contribution
(the skew scattering \citep{SMIT1955Phys} and side jump \citep{Berger1970PRB})
of conventional AHE \citep{Sinitsyn2007PRB,Ado2015,Nunner2008PRL,YangSY2011PRB,LuHZ2013PRB,ShanWY2013PRB,Burkov2014PRL,Ado2016PRL,Konig2017PRL,ZhangJX2023PRB},
which sometimes becomes comparable to or even predominant over the
intrinsic contribution in kagome metals \citep{YangSY2020SA,Yu2021PRB,Zheng2023NC}.
Current investigations of IPAHE mainly focused on the intrinsic part
due to the Berry curvature of energy bands, however, studies of the
extrinsic part due to the disorder scattering remain limited \citep{WangCM2023PRB}.
In particular, how inevitable spin-dependent scattering from magnetic
disorder affects the IPAHE is crucial for understanding the Hall transport
experiments and to develop the energy-efficient electronic devices
in magnetic materials. In realistic magnetic materials, electrons
are scattered not only by scalar (nonmagnetic) impurities but also
by spin dependent potentials originating from local moments and spin
fluctuations.

In this work, we systematically investigate the impact of different
universality classes of disorder on the IPAHE using the Kubo formula
with two-dimensional (2D) massive Dirac model with a hexagonal-warping
term. We find that each universal class of scattering produces distinct
behaviors of the extrinsic parts (the skew scattering and side jump)
of AHE. We recover the previous results of AHE in conventional massive
Dirac fermions in the presence of three universality classes of disorder
and in Dirac fermions with the warping term with nonmagnetic impurities.
In addition, we calculate the in-plane magnetoresistance that enables
us to better understand previous experimental results. 

The rest of the work is organized as follows. In Sec. II, we briefly
introduce the Dirac model with the warping term under general Zeeman
fields and calculate the intrinsic anomalous Hall conductivity. In
Sec. III, we discuss the basic physics of three universal classes
of disorder scattering. Section IV presents the main results of the
extrinsic AHC. In Sec. V, we calculate the in-plane magnetoresistance
and make some comparisons with previous results. Finally, we draw
some conclusions.

\section{model for Dirac fermions with warping term }

\begin{figure}
\includegraphics[width=8.5cm]{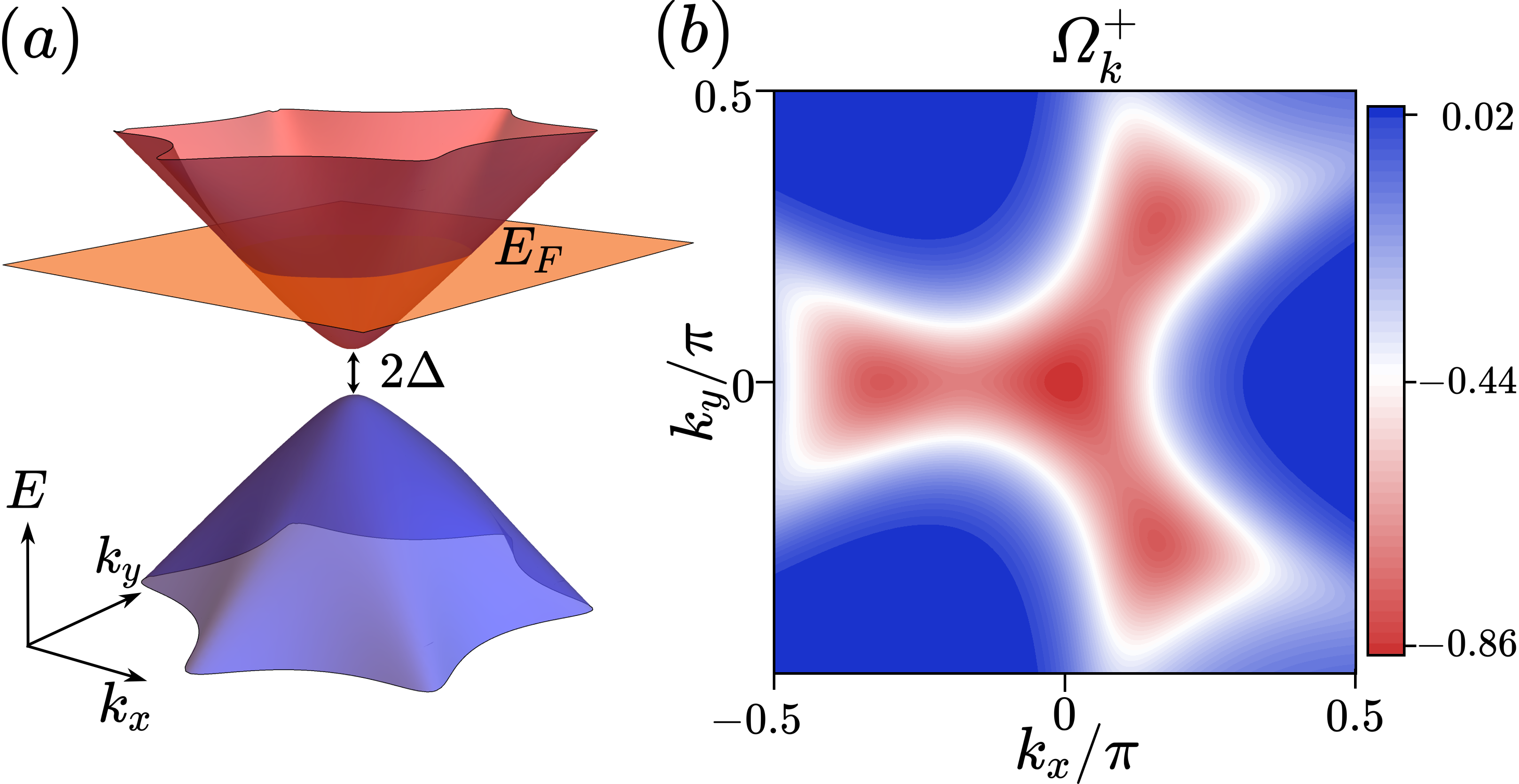}
\begin{turnpage}
\caption{\label{fig:Ek} (a) Schematic of the energy band of the Dirac surface
states of the topological insulator with the hexagonal warping. The
Fermi energy lies in the conduction band. (b) The distribution of
the Berry curvature of the conduction band with only out-of-plane
magnetization $\mathcal{M}_{z}$ in the momentum space, the parameters
are given as $\lambda=0.1\ \mathrm{eV\ nm^{3}},v=0.2\ \mathrm{eV\ nm},\ \mathcal{M}_{z}=0.15\ \mathrm{eV}$.}
\end{turnpage}

\end{figure}

We consider the typical two-band model supporting the IPAHE and its
quantized counterpart, which can well describe the surface states
of a topological insulator with the hexagonal warping term \citep{LiuX2013PRL}
\begin{equation}
H_{0}\left(\bm{k}\right)=v(k_{y}\sigma_{x}-k_{x}\sigma_{y})+\lambda_{k}\sigma_{z}+\boldsymbol{\mathcal{M}}\cdot\hat{\boldsymbol{\sigma}},\label{eq:H}
\end{equation}
where $\hat{\boldsymbol{\sigma}}=(\sigma_{x},\sigma_{y},\sigma_{z})$
are the Pauli matrices acting in the spin space, $v$ is the Dirac
velocity, and $\boldsymbol{k}=(k_{x},k_{y})$ is the 2D wave vector.
The first term is the Rashba-type spin-momentum locking, the second
term $\lambda_{k}=\lambda k_{x}(k_{x}^{2}-3k_{y}^{2})$ is the generic
hexagonal warping term \citep{FuL2009PRL}, and the third term is
the Zeeman coupling or the magnetization energy $\boldsymbol{\mathcal{M}}=\left(\mathcal{\boldsymbol{M}}_{\parallel},\mathcal{M}_{z}\right)$
where $\mathcal{\boldsymbol{M}}_{\parallel}=(\mathcal{M}_{\parallel}\cos\theta,\mathcal{M}_{\parallel}\sin\theta)=g\mu_{B}\boldsymbol{B}/2$
is the in-plane component and $\mathcal{M}_{z}$ is the out-of-plane
component. $g$ is the effective $g$ factor and $\mu_{B}$ is the
Bohr magneton. The energy dispersion is of the form
\begin{equation}
\varepsilon_{\boldsymbol{k}}^{\pm}=\pm\sqrt{v^{2}q^{2}+\left(\lambda_{k}+\mathcal{M}_{z}\right)^{2}},
\end{equation}
and is depicted in Fig.\ref{fig:Ek}(a), with $\boldsymbol{q}=(q_{x},q_{y})=\left[(-vk_{y}+\mathcal{M}_{x})/v,(vk_{x}+\mathcal{M}_{y})/v\right]$.
The corresponding eigenstates are $\psi_{\boldsymbol{k}}^{\pm}(\boldsymbol{r})=\left|u_{\boldsymbol{k}}^{\pm}\right\rangle e^{i\boldsymbol{k}\cdot\boldsymbol{r}}$
with
\begin{equation}
\left|u_{\boldsymbol{k}}^{+}\right\rangle =\left(\begin{array}{c}
\cos\frac{\Theta_{\boldsymbol{k}}}{2}\\
\sin\frac{\Theta_{\boldsymbol{k}}}{2}e^{i\phi_{\boldsymbol{k}}}
\end{array}\right),\left|u_{\boldsymbol{k}}^{-}\right\rangle =\left(\begin{array}{c}
\sin\frac{\Theta_{\boldsymbol{k}}}{2}\\
-\cos\frac{\Theta_{\boldsymbol{k}}}{2}e^{i\phi_{\boldsymbol{k}}}
\end{array}\right),
\end{equation}
where $\Theta_{\boldsymbol{k}}=\tan^{-1}\left.vq\right/\left(\lambda(k_{x}^{3}+k_{x}k_{y}^{2})+\mathcal{M}_{z}\right)$,
$\phi=\tan^{-1}\left.q_{y}\right/q_{x}$. The Zeeman coupling caused
by the in-plane field shifts the Dirac point away from the point $\boldsymbol{k}\equiv0$
to $\boldsymbol{k}'\equiv\hat{\boldsymbol{z}}\times\mathcal{\boldsymbol{M}}_{\parallel}/v$
\citep{FuL2009PRL}, and the hexagonal warping term then opens a small
gap $2\Delta_{1}\equiv2\lambda k'{}^{3}\sin3\theta$ at the Dirac
point \citep{SunS2022PRB}. Similarly, the external magnetic field
in the z-direction will also open the energy gap $2\Delta_{2}\equiv2\mathcal{M}_{z}$
of the system. The effective gap becomes $2\Delta\approx2(\Delta_{1}+\Delta_{2})$,
leading to a sizable AHE. 

It is straightforward to evaluate the Berry curvature as
\begin{equation}
\varOmega_{\boldsymbol{k}}^{\pm}=\frac{\pm v}{2\varepsilon_{\boldsymbol{k}}^{3}}\left[v\left(2\lambda_{k}-\mathcal{M}_{z}\right)+3\lambda(\mathcal{M}_{y}(k_{x}^{2}-k_{y}^{2})+2\mathcal{M}_{x}k_{x}k_{y})\right].
\end{equation}
The integration of $\varOmega_{\boldsymbol{k}}^{\pm}$ over the Brillouin
zone gives rise to the intrinsic anomalous Hall conductivity to leading
order in the warping parameter (details can be found in the Supplemental
Material \citep{SMs})
\begin{align}
\sigma_{xy}^{in} & =-\frac{e^{2}}{h}\left(\frac{\mathcal{M}_{z}}{2E_{F}}+\frac{\lambda\mathcal{M}_{\parallel}^{3}\sin3\theta}{2v^{3}E_{F}}\right).
\end{align}
where $\theta$ is the angle between the in-plane magnetic field or
magnetization and the $x$ axis. We show the distribution of the Berry
curvature of the conduction band in the momentum space in Fig.\ref{fig:Ek}(b).
The first contribution comes from the out-of-plane magnetic field
or magnetization, while the second one is induced by the in-plane
magnetic field that breaks the combination of time-reversal symmetry
and mirror symmetry. When the Fermi level lies in the band gap, the
first term reduces to the known half-quantized Hall conductivity \citep{shen2013TIDirac}.
Note that the higher-order term corresponds to the magnetic octupole
contribution and reflects the anisotropic Fermi surface due to the
warping term with $C_{3v}$ symmetry. 

\section{Universal classes of disorder scattering}

To simulate the spin-dependent scattering in magnetic materials, we
consider the general form of a random disorder potential for carriers
with spin (or pseudo spin) degrees of freedom 
\begin{equation}
\hat{V}_{dis}(\boldsymbol{r})=\sum_{i}\left(V_{0}\hat{\sigma}_{0}+\boldsymbol{V}\cdot\hat{\boldsymbol{\sigma}}\right)\delta(\boldsymbol{r}-\boldsymbol{R}_{i}),
\end{equation}
where $\boldsymbol{R}_{i}$ $(i=1,2,\ldots)$ labels positions of
randomly distributed scattering centers, and $V_{0}$ denotes the
scalar (nonmagnetic) component of the impurity potential, while $\boldsymbol{V}=(V_{x},V_{y},V_{z})$
correspond to the spin-dependent (magnetic) ones.

We focus on three different types of disorder. First, in ferromagnetic
materials, normal (nonmagnetic) impurity scattering and phonon scattering
belong to the class A ($V_{0}\hat{\sigma}_{0}$ type impurity). Second,
the class B ($V_{z}\hat{\sigma}_{z}$ type impurity) refers to scattering
by magnetic impurities that conserve the $z$ component of the carrier
spin. Third, the scattering processes of the class C ($V_{x}\hat{\sigma}_{x}$
and $V_{y}\hat{\sigma}_{y}$ type impurities) are due to the spin-flipping
scattering ($s-d$ interaction) by in-plane random magnetic impurities
or the magnetic fluctuations of in-plane magnetic order \citep{YangSY2011PRB,LiWR2024jpcm}
\begin{align}
H_{int} & =-J\int dr\boldsymbol{S}(r)\cdot\hat{\boldsymbol{\sigma}}(r),\nonumber \\
 & =-\frac{J}{2}\int dr\left[\hat{\sigma}_{+}S_{-}+\hat{\sigma}_{-}S_{+}+2\hat{\sigma}_{z}S_{z}\right].\label{eq:Himp}
\end{align}

We assume that the statistical average of the disorder potential is
zero since any nonzero value only shifts the origin of total energy,
and that the second-order spatial correlation only depends on the
difference in positions within the Gaussian approximation. Therefore,
for the different types of impurities listed above and the calculations
presented in the following, we have the zero average value $\left\langle V_{dis,\boldsymbol{k},\boldsymbol{k}^{\prime}}^{\eta\eta^{\prime}}\right\rangle _{imp}=0$,
where $\eta,\eta^{\prime}=\pm$ indicates the matrix element between
the eigenstates $\eta$ and $\eta^{\prime}$, and the angular brackets
$\left\langle \ldots\right\rangle _{imp}$ denote the disorder average
\citep{bruus2004mbqt}. We then have the Gaussian correlations between
disorders,
\begin{align}
 & \left\langle V_{\boldsymbol{k},\boldsymbol{k}^{\prime}}^{l,\eta\eta^{\prime}}V_{\boldsymbol{k}^{\prime},\boldsymbol{k}}^{m,\eta^{\prime}\eta}\right\rangle _{imp}\nonumber \\
= & \frac{n_{i}u_{0}^{2}}{V}\left\langle u_{\boldsymbol{k}}^{\eta}\left|\sigma_{l}\right|u_{\boldsymbol{k}^{\prime}}^{\eta^{\prime}}\right\rangle \left\langle u_{\boldsymbol{k}^{\prime}}^{\eta^{\prime}}\left|\sigma_{m}\right|u_{\boldsymbol{k}}^{\eta}\right\rangle ,
\end{align}
where $l,m=x,y,z$; $n_{i}$ is the impurity concentration, and $u_{0}$
is the disorder strength in Gaussian approximation. Furthermore, to
evaluate the impacts of the skew scattering due to anisotropic part
of the scattering rate, we need to take into account at least the
third-order disorder non-Gaussian correlations,
\begin{align}
 & \left\langle V_{\boldsymbol{k},\boldsymbol{k}^{\prime}}^{l,\eta\eta^{\prime}}V_{\boldsymbol{k}^{\prime},\boldsymbol{k}^{\prime\prime}}^{m,\eta^{\prime}\eta^{\prime\prime}}V_{\boldsymbol{k}^{\prime\prime},\boldsymbol{k}}^{n,\eta^{\prime\prime}\eta}\right\rangle _{imp}\nonumber \\
 & =\frac{n_{i}u_{1}^{3}}{V^{2}}\left\langle u_{\boldsymbol{k}}^{\eta}\left|\sigma_{l}\right|u_{\boldsymbol{k}}^{\eta^{\prime}}\right\rangle \left\langle u_{\boldsymbol{k}}^{\eta^{\prime}}\left|\sigma_{m}\right|u_{\boldsymbol{k}^{\prime\prime}}^{\eta^{\prime\prime}}\right\rangle \left\langle u_{\boldsymbol{k}^{\prime\prime}}^{\eta^{\prime\prime}}\left|\sigma_{n}\right|u_{\boldsymbol{k}}^{\eta}\right\rangle ,
\end{align}
where $u_{1}$ is the disorder strength in non-Gaussian approximation.
Next, we would separately calculate the anomalous Hall conductivity
caused by three impurity scattering classes. 
\begin{figure}
\includegraphics[width=8.5cm]{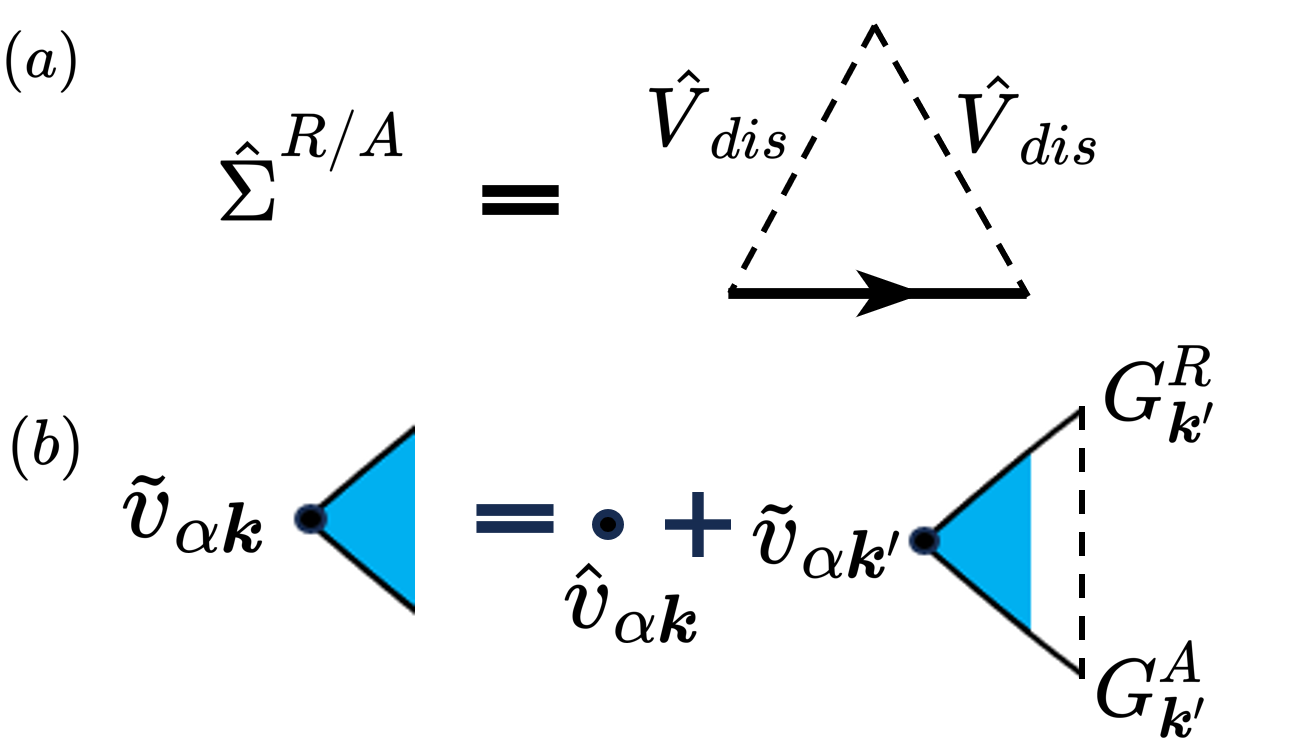}
\begin{turnpage}
\caption{\label{fig:ladder diagrams} (a) The electron self-energy due to the
impurity scattering in the first Born approximation. (b) The vertex
correction of the velocity operator $\tilde{v}_{\alpha\boldsymbol{k}}$
for calculating the electric conductivity $\sigma_{\alpha\beta}$.}
\end{turnpage}

\end{figure}

\section{extrinsic anomalous Hall effect }

\subsection{Kubo formula}

The Kubo formula provides us with a systematic way to calculate the
anomalous Hall conductivity in the weak scattering regime and reveals
some fundamental features of AHE in magnetic materials \citep{Streda1982JPC},
\begin{align}
\sigma_{xy}^{\textrm{total}} & =\sigma_{xy}^{\mathrm{I}}+\sigma_{H,xy}^{\mathrm{II}},\\
\sigma_{xy}^{\mathrm{I}} & =\frac{e^{2}\hbar}{2\pi}\sum_{\boldsymbol{k}}\mathrm{Tr}\left\langle \hat{v}_{x\boldsymbol{k}}G_{\boldsymbol{k}}^{R}\hat{v}_{y\boldsymbol{k}}G_{\boldsymbol{k}}^{A}\right\rangle _{imp},\\
\sigma_{xy}^{\mathrm{\mathrm{II}}} & =-\sigma_{yx}^{\mathrm{\mathrm{II}}}=ec\left.\frac{\partial n(E)}{\partial B}\right|_{E=E_{F},B=0}.
\end{align}
The first term $\sigma_{xy}^{\mathrm{I}}$ describes the contribution
of the electrons in the conduction band near the Fermi surface while
$\sigma_{xy}^{\mathrm{II}}$ accounts for the contribution to the
entire Fermi sea. Note that $\sigma_{xy}^{\mathrm{II}}$ plays a key
role in understanding the topological nature of the integer quantum
Hall effect from the point of view of thermodynamics. 

Throughout this work we focus on the metallic weak-scattering regime
and adopt a Gaussian white-noise disorder model. Technically, we evaluate
the conductivity within the Kubo-Středa formalism using disorder-averaged
Green’s functions and ladder-type vertex corrections, i.e., the noncrossing
approximation (NCA). To proceed, we calculate the averaged Green's
function by solving the Dyson equation in the first Born approximation
as shown in Fig.~\ref{fig:ladder diagrams}(a), where the full retarded/advanced
Green’s functions are given by
\begin{equation}
G^{R/A}=\frac{1}{E_{F}-\hat{H}\pm i\varGamma},
\end{equation}
where $\varGamma=\Gamma_{0}\sigma_{0}+\Gamma_{x}\sigma_{x}+\Gamma_{y}\sigma_{y}+\Gamma_{z}\sigma_{z}$
is the imaginary part of the self energy $\Sigma^{R/A}$. The self
energy in the first Born approximation is
\begin{align}
\Sigma^{R/A} & =\sum_{\boldsymbol{k}}\left\langle V_{dis}G_{\boldsymbol{k}}^{R/A}V_{dis}\right\rangle _{imp}.
\end{align}
Meanwhile, the vertex correction in the ladder approximation {[}in
Fig. \ref{fig:ladder diagrams}(b){]} \citep{Sinitsyn2007PRB} at
the Fermi energy is given by
\begin{align}
\tilde{v}_{x/y\boldsymbol{k}} & =\hat{v}_{x/y\boldsymbol{k}}+\sum_{\boldsymbol{k}'}\left\langle V_{dis}G_{\boldsymbol{k}'}^{A/R}\tilde{v}_{x/y\boldsymbol{k}'}G_{\boldsymbol{k}'}^{R/A}V_{dis}\right\rangle _{imp},
\end{align}
which determines the corrected vertex function of $\tilde{v}_{\alpha\boldsymbol{k}}$
from the bare velocity operator $\hat{v}_{\alpha\boldsymbol{k}}$
with $\alpha=x,y$.

For electron conduction and up to the first order of $\lambda$ at
zero temperature and zeroth order of impurity concentration $n_{i}$,
the total Hall conductivity can be calculated analytically in the
spin basis or eigenstate basis (details are given in Supplemental
Material \citep{SMs}).
\begin{align}
\sigma_{xy} & =\left(\sigma_{xy}^{in}+\sigma_{xy}^{sj,2}+\sigma_{xy}^{sk,4}\right)+\sigma_{xy}^{sk,3}.
\end{align}
In this work, we are mostly interested in the Hall conductivity $\sigma_{xy}$
of the order $n_{i}^{0}$ ($\sigma_{xy}^{0}=\sigma_{xy}^{in}+\sigma_{xy}^{sj,2}+\sigma_{xy}^{sk,4}$)
and of the order $n_{i}^{-1}$ ($\sigma_{xy}^{-1}=\sigma_{xy}^{sk,3}$).
Here we follow the conventions in \citep{Sinitsyn2007PRB}. At low
temperatures, the thermal smearing of the Fermi surface and temperature
dependence of scattering rates need to be considered.

\subsection{Side jump part}

\begin{figure}
\includegraphics[width=8.6cm]{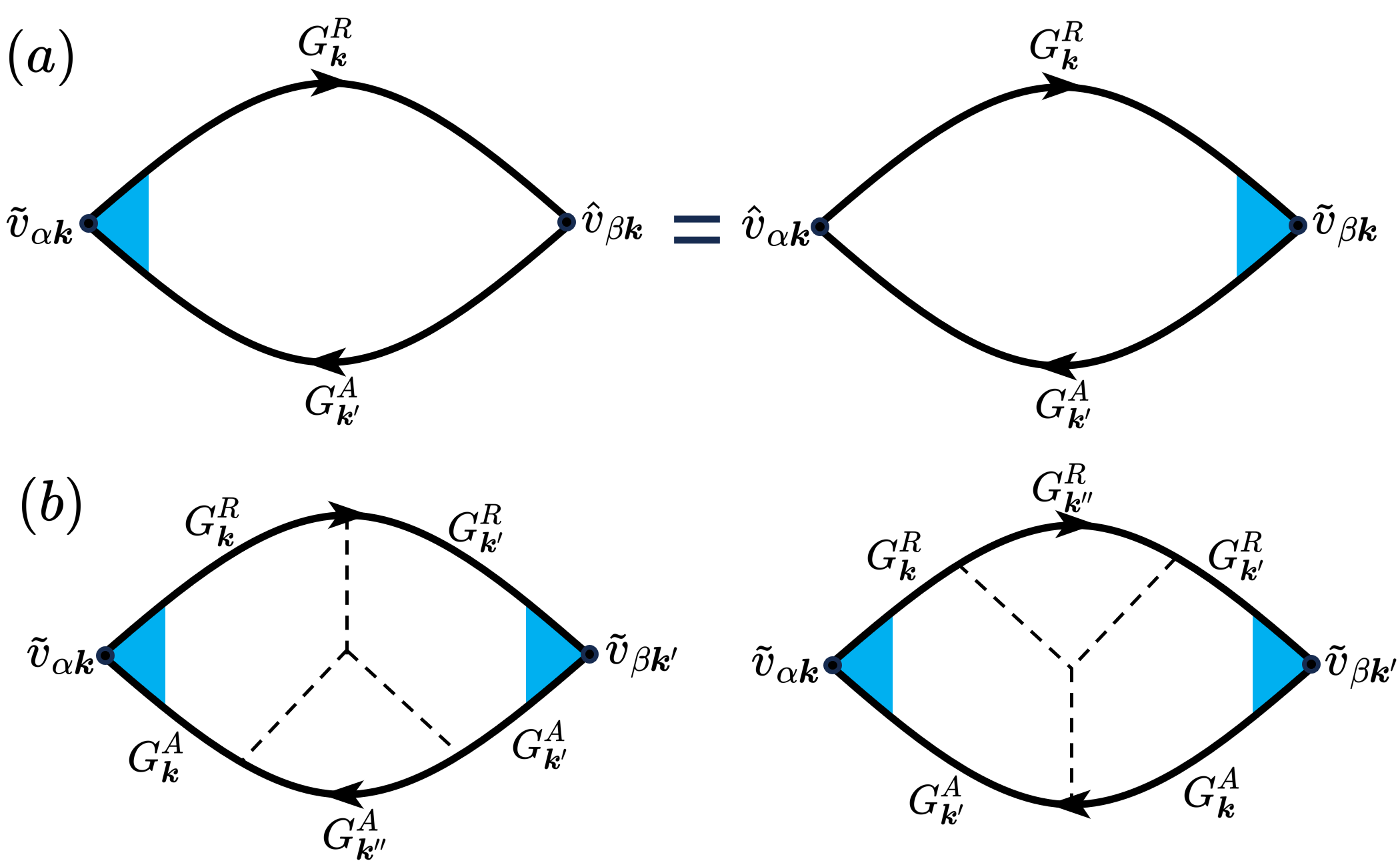}
\begin{turnpage}
\caption{\label{fig:second order sj-1} The Feynman diagrams for calculating
the electric conductivity tensor $\sigma_{\alpha\beta}$. (a) Total
conductivity of order $n_{dis}^{0}$, (b) Total conductivity of order
$n_{dis}^{-1}$.}
\end{turnpage}

\end{figure}

We first calculate the contribution of the side jump through the Kubo
formula approach. In the semiclassical picture, the side jump we define
here consists of three components: the coordinate shift, a correction
of the distribution function, and some higher-order scattering processes
(the intrinsic skew scattering) \citep{Sinitsyn2007PRB}. Figure \ref{fig:second order sj-1}(a)
shows a set of diagrams that contribute to the intrinsic and side
jump in the spin basis. The contribution to Hall conductivity from
the second-order side jump contribution for each scattering class
is $\sigma_{xy}^{sj,2}$, and the contribution of the fourth-order
scattering processes (intrinsic skew scattering) is $\sigma_{xy}^{sk,4}$.
Then the total side jump contribution to anomalous Hall conductivity
is given by $\sigma_{xy}^{sj}=\sigma_{xy}^{sj,2}+\sigma_{xy}^{sk,4}$:
\begin{align}
\mathrm{Class}\  & A:\nonumber \\
\sigma_{xy,V_{0}}^{sj} & =\frac{e^{2}}{h}\left(\frac{\mathcal{M}_{z}}{2E_{F}}-\frac{4E_{F}\mathcal{M}_{z}\left(E_{F}^{2}+\mathcal{M}_{z}^{2}\right)}{\left(E_{F}^{2}+3\mathcal{M}_{z}^{2}\right)^{2}}\right)\nonumber \\
 & +\left(1-\frac{8E_{F}^{2}\left(E_{F}^{4}-6E_{F}^{2}\mathcal{M}_{z}^{2}-3\mathcal{M}_{z}^{4}\right)}{\left(E_{F}^{2}+3\mathcal{M}_{z}^{2}\right)^{3}}\right)\nonumber \\
 & \times\frac{e^{2}}{h}\frac{\lambda\mathcal{M}_{\parallel}^{3}\sin3\theta}{2v^{3}E_{F}}.\label{eq:SJV0}\\
\mathrm{Class}\  & B:\nonumber \\
\sigma_{xy,V_{z}}^{sj} & =\frac{e^{2}}{h}\left(\frac{\mathcal{M}_{z}}{2E_{F}}-\frac{4E_{F}\mathcal{M}_{z}\left(E_{F}^{2}+\mathcal{M}_{z}^{2}\right)}{\left(3E_{F}^{2}+\mathcal{M}_{z}^{2}\right)^{2}}\right)\nonumber \\
 & +\left(1-\frac{8E_{F}^{2}\left(3E_{F}^{4}+6E_{F}^{2}\mathcal{M}_{z}^{2}-\mathcal{M}_{z}^{4}\right)}{\left(3E_{F}^{2}+\mathcal{M}_{z}^{2}\right)^{3}}\right)\nonumber \\
 & \times\frac{e^{2}}{h}\frac{\lambda\mathcal{M}_{\parallel}^{3}\sin3\theta}{2v^{3}E_{F}}.\\
\mathrm{Class}\  & C:\nonumber \\
\sigma_{xy,V_{x}}^{sj} & =\frac{e^{2}}{h}\left(\frac{\mathcal{M}_{z}}{2E_{F}}-\frac{2\lambda E_{F}\mathcal{M_{\parallel}}\sin\theta}{v^{3}}+\frac{\lambda\mathcal{M}_{\parallel}^{3}\sin3\theta}{2v^{3}E_{F}}\right).\label{SJC1}\\
\sigma_{xy,V_{y}}^{sj} & =\frac{e^{2}}{h}\left(\frac{\mathcal{M}_{z}}{2E_{F}}+\frac{2\lambda E_{F}\mathcal{M}_{\parallel}\sin\theta}{v^{3}}+\frac{\lambda\mathcal{M}_{\parallel}^{3}\sin3\theta}{2v^{3}E_{F}}\right).\label{SJC2}
\end{align}
There are several features of the side jump contributions. First,
for all of the three classes, they are independent of disorder density
and scattering strength, similar to the pure intrinsic AHE. Second,
the side jump contributions contain two parts from the out-plane magnetization
$\mathcal{M}_{z}$ and in-plane one $\mathcal{M}$ and exhibit distinct
dependence of magnetization. It may originate from the different self-energy
and $k$ dependence of the scattering vertices. Third, the threefold
rotational symmetry is preserved in the contribution from the in-plane
magnetization in the first two classes (A and B) but gets broken in
class C. Note that in the class C, by combining with intrinsic one,
it leaves the only part with a $2\pi$ period. 

\subsection{Skew scattering part}

We turn to calculate the skew scattering contribution for each scattering
class. Fig. \ref{fig:second order sj-1}(b) show a set of diagrams
that contribute to the skew scattering in the spin basis. The skew
scattering contribution comes from the asymmetric part of the scattering
rates for higher order scattering processes \citep{SMIT1955Phys},
and its contribution to anomalous Hall conductivity depends on disorder
density and scattering strength:
\begin{align}
\mathrm{Class}\ \mathrm{A}: & \ \sigma_{xy,V_{0}}^{sk}=-\frac{e^{2}}{h}\frac{u_{1}^{3}}{n_{i}u_{0}^{4}}\mathcal{M}_{z}\frac{\left(E_{F}^{2}-\mathcal{M}_{z}^{2}\right)^{2}}{\left(E_{F}^{2}+3\mathcal{M}_{z}^{2}\right)^{2}}\nonumber \\
- & \frac{e^{2}}{h}\frac{u_{1}^{3}}{n_{i}u_{0}^{4}}\frac{\lambda\mathcal{M}_{\parallel}^{3}\sin3\theta}{v^{3}}\frac{\left(E_{F}^{2}-\mathcal{M}_{z}^{2}\right)}{\left(E_{F}^{2}+3\mathcal{M}_{z}^{2}\right)^{3}}\nonumber \\
\times & \left(E_{F}^{4}-14E_{F}^{2}\mathcal{M}_{z}^{2}-3\mathcal{M}_{z}^{4}\right).\label{SSV0}\\
\mathrm{Class}\ B: & \ \sigma_{xy,V_{z}}^{sk}=\frac{e^{2}}{h}\frac{u_{1}^{3}}{n_{i}u_{0}^{4}}E_{F}\frac{\left(E_{F}^{2}-\mathcal{M}_{z}^{2}\right){}^{2}}{\left(3E_{F}^{2}+\mathcal{M}_{z}^{2}\right)^{2}}\nonumber \\
- & \frac{e^{2}}{h}\frac{u_{1}^{3}}{n_{i}u_{0}^{4}}\frac{16\lambda\mathcal{M}_{\parallel}^{3}\sin3\theta}{v^{3}}\frac{\mathcal{M}_{z}E_{F}^{3}\left(E_{F}^{2}-\mathcal{M}_{z}^{2}\right)}{\left(3E_{F}^{2}+\mathcal{M}_{z}^{2}\right)^{3}}.\label{SSCB}\\
\mathrm{Class}\ C: & \ \sigma_{xy,V_{x}}^{sk}=\frac{e^{2}}{h}\frac{u_{1}^{3}}{n_{i}u_{0}^{4}}\frac{\lambda E_{F}\mathcal{M}_{\parallel}^{2}\sin2\theta}{v^{3}}.\label{SSC1}\\
 & \sigma_{xy,V_{y}}^{sk}=\frac{e^{2}}{h}\frac{u_{1}^{3}}{n_{i}u_{0}^{4}}\frac{\lambda E_{F}\mathcal{M}_{\parallel}^{2}\cos2\theta}{v^{3}}.\label{SSC2}
\end{align}
The skew scattering contribution of each class impurity is dependent
on the $r=(n_{i}u_{1}^{3})^{2/3}/(n_{i}u_{0}^{2})$ and is inversely
proportional to the impurity density. In high mobility conductors,
the skew scattering part could dominate over both the intrinsic and
side jump parts, such as the kagome metals \citep{Zheng2023NC}. Similar
to the results in \citep{YangSY2011PRB}, the skew scattering corrections
in class A and class B respect the $C_{3v}$ symmetry. However, in
class C, the skew scattering contribution becomes quadratic in $\mathcal{M}_{\parallel}$
with a $\pi$ period, indicating the breaking of $C_{3v}$ symmetry. 

Let us understand the unusual angle dependence of extrinsic AHE due
to magnetic scattering. In the presence of magnetic impurity scattering,
the Onsager relation could be 
\begin{align}
\sigma_{\alpha\beta}^{H}(\boldsymbol{\mathcal{M}},\boldsymbol{u}) & =\sigma_{\beta\alpha}^{H}(-\boldsymbol{\mathcal{M}},-\boldsymbol{u}),\label{eq:Onsager relation}
\end{align}
where $\boldsymbol{\mathcal{M}}$ is the direction vector of the magnetic
field or magnetization and $\boldsymbol{u}$ denotes the direction
vector of magnetic impurity. The Onsager relation excludes all the
even order terms of the total order of the magnetic impurity strength
and magnetic fields. Since the skew scattering is proportional to
the odd order of magnetic impurity strength, the Hall conductivity
should be proportional to the zero order and even order of the total
magnetic field, such as $u_{1}^{3}/n_{i}u_{0}^{4}$ and $\left(u_{1}^{3}/n_{i}u_{0}^{4}\right)\mathcal{M}_{\parallel}^{2}\sin2\theta$.
Note that the in-plane magnetic scattering (class C) lowers the crystal
symmetry and results in a side jump contribution proportional to $\mathcal{M}_{\parallel}\sin(\theta)$
in Eqs. $\ref{SJC1}$ and $\ref{SJC2}$.

\subsection{Total anomalous Hall conductivity}

We collect both the intrinsic and extrinsic parts and reach the total
anomalous Hall conductivity for each universality class $\sigma_{xy}=\sigma_{xy}^{in}+\sigma_{xy}^{sj}+\sigma_{xy}^{sk}$:
\begin{align}
\mathrm{Class}\  & A:\nonumber \\
\sigma_{xy,V_{0}} & =-\frac{e^{2}}{h}\frac{4E_{F}\mathcal{M}_{z}\left(E_{F}^{2}+\mathcal{M}_{z}^{2}\right)}{\left(E_{F}^{2}+3\mathcal{M}_{z}^{2}\right)^{2}}-\frac{e^{2}}{h}\frac{4\lambda\mathcal{M}_{\parallel}^{3}\sin3\theta}{v^{3}E_{F}}\nonumber \\
 & \times\frac{E_{F}^{2}\left(E_{F}^{4}-6E_{F}^{2}\mathcal{M}_{z}^{2}-3\mathcal{M}_{z}^{4}\right)}{\left(E_{F}^{2}+3\mathcal{M}_{z}^{2}\right)^{3}}\nonumber \\
 & -\frac{e^{2}}{h}\frac{u_{1}^{3}}{n_{i}u_{0}^{4}}\frac{\left(E_{F}^{2}-\mathcal{M}_{z}^{2}\right)}{\left(E_{F}^{2}+3\mathcal{M}_{z}^{2}\right)^{2}}\left[\mathcal{M}_{z}\left(E_{F}^{2}-\mathcal{M}_{z}^{2}\right)\right.\nonumber \\
 & \left.+\frac{\lambda\mathcal{M}_{\parallel}^{3}\sin3\theta}{v^{3}}\frac{\left(E_{F}^{4}-14E_{F}^{2}\mathcal{M}_{z}^{2}-3\mathcal{M}_{z}^{4}\right)}{\left(E_{F}^{2}+3\mathcal{M}_{z}^{2}\right)}\right].\label{eq:V0}\\
\mathrm{Class}\  & B:\nonumber \\
\sigma_{xy,V_{z}} & =-\frac{e^{2}}{h}\frac{4E_{F}\mathcal{M}_{z}\left(E_{F}^{2}+\mathcal{M}_{z}^{2}\right)}{\left(3E_{F}^{2}+\mathcal{M}_{z}^{2}\right)^{2}}-\frac{e^{2}}{h}\frac{4\lambda\mathcal{M}_{\parallel}^{3}\sin3\theta}{v^{3}E_{F}}\nonumber \\
 & \times\frac{E_{F}^{2}\left(3E_{F}^{4}+6E_{F}^{2}\mathcal{M}_{z}^{2}-\mathcal{M}_{z}^{4}\right)}{\left(3E_{F}^{2}+\mathcal{M}_{z}^{2}\right)^{3}}\nonumber \\
 & +\frac{e^{2}}{h}\frac{u_{1}^{3}}{n_{i}u_{0}^{4}}\frac{\left(E_{F}^{2}-\mathcal{M}_{z}^{2}\right)}{\left(3E_{F}^{2}+\mathcal{M}_{z}^{2}\right)^{2}}\left[E_{F}\left(E_{F}^{2}-\mathcal{M}_{z}^{2}\right)\right.\nonumber \\
 & \left.-\frac{16\lambda\mathcal{M}_{\parallel}^{3}\sin3\theta}{v^{3}}\frac{\mathcal{M}_{z}E_{F}^{3}}{\left(3E_{F}^{2}+\mathcal{M}_{z}^{2}\right)}\right].\\
\mathrm{Class}\  & C:\nonumber \\
\sigma_{xy,V_{x}} & =\frac{e^{2}\lambda E_{F}}{h}\left(-\frac{2\mathcal{M_{\parallel}}\sin\theta}{v^{3}}+\frac{u_{1}^{3}}{n_{i}u_{0}^{4}}\frac{\mathcal{M}_{\parallel}^{2}\sin2\theta}{v^{3}}\right).\label{eq:ipaheX}\\
\sigma_{xy,V_{y}} & =\frac{e^{2}\lambda E_{F}}{h}\left(\frac{2\mathcal{M}_{\parallel}\sin\theta}{v^{3}}+\frac{u_{1}^{3}}{n_{i}u_{0}^{4}}\frac{\mathcal{M}_{\parallel}^{2}\cos2\theta}{v^{3}}\right).\label{eq: ipahe}
\end{align}
Let us summarize several salient features here. First of all, since
the scattering in class A and class B respect the $C_{3v}$ symmetry,
the anomalous Hall conductivity merely has the threefold symmetric
part. Second, in class C, the in-plane spin flipping scattering further
lowers the symmetry and thus give rise to the nonzero $\pi$ and $2\pi$
periodic anomalous Hall conductivity, which is distinct from the isotropic
massive Dirac fermions with vanishing skew scattering contribution
\citep{YangSY2011PRB}. Third, in class B, there exists an unusual
skew scattering contribution that is independent of the in-plane and
out-of-plane magnetizations {[}$\sigma_{xy,V_{z}}^{sk}\approx\frac{e^{2}}{h}\frac{u_{1}^{3}E_{F}}{9n_{i}u_{0}^{4}}$
from the first line of Eq. $\left(\ref{SSCB}\right)$ in the limit
of vanishing $\mathcal{M}_{z}${]}. It may originate from the gap
renormalization induced by the disorders $V_{z}$. Note that in the
limit of $\lambda\shortrightarrow0$ or $\mathcal{M}_{z}\shortrightarrow0$,
our results could recover the previous ones \citep{YangSY2011PRB,WangCM2023PRB}.
In sum, the crystal symmetry such as $C_{3v}$ here greatly enrich
the features of AHE in the presence of different classes of disorder
scattering. 

To illustrate the impacts of in-plane impurity concentration on IPAHE,
we calculate the conductivity in the presence of impurities with ratio
of $nV_{x}+(1-n)V_{y}$, $\left(0\leq n\leq1\right)$. The resulting
Hall conductivity interpolates smoothly between the pure $nV_{x}$
{[}$n=1$, Eq.\eqref{eq:ipaheX}{]} and pure $V_{y}$ {[}$n=0$, Eq.\eqref{eq: ipahe}{]}
limits:
\begin{align}
\sigma_{xy,V_{x};V_{y}} & =\frac{2e^{2}\lambda\mathcal{M}_{\parallel}E_{F}}{hv^{3}}\frac{(1-2n)\sin\theta+2(n-1)n\cos\theta}{2n(n-1)+1},\nonumber \\
- & \frac{u_{1}^{3}}{n_{i}u_{0}^{4}}\frac{e^{2}\lambda\mathcal{M}_{\parallel}^{2}E_{F}}{hv^{3}}\frac{(n-1)\cos2\theta-n\sin2\theta}{2n(n-1)+1},
\end{align}
which shows that both the amplitude and the angular dependence of
the IPAHE are sensitive to the relative weights of different in-plane
impurity components. For the in-plane impurities with a fixed weight
$n$, as long as this weight does not vary upon rotating the external
magnetic field (i.e., $n$ is independent of $\theta$), the in-plane
Hall conductivity retains the well-defined $\pi$ and $2\pi$ periodic
components.

In our work, the “spin-dependent disorder” terms are parameterized
by the spin-conserving and spin-flip components of the impurity potential
(Eq.\eqref{eq:Himp}). Microscopically, such terms may originate from
localized moments associated with defects/dopants or from slowly varying
magnetic textures/fluctuations, depending on the material properties.
Therefore, our treatment approach is phenomenological, aiming to classify
the role of disorder of different spin structures in IPAHE. Note that
if the relevant moments behave as freely fluctuating para-magnetic
impurities, a sufficiently strong in-plane magnetic field could polarize
them {[}$n=n(\theta)${]} and suppress the spin-flip processes, thereby
continuously diminishing the $\pi$-periodic component associated
with the spin-flip class. Taking account of such field-dependent polarization
would cause a field- and angle-dependent weight $n$, which is beyond
the scope of the present work.

\subsection{The in-plane Hall conductivity of $\mathrm{X}$ and $\Psi$ diagrams}

It is known that, for the conventional AHE of 2D massive Dirac fermions,
the crossed diagrams could contribute the same order for close impurity
pairs (the so-called $\mathrm{X}$ and $\Psi$ diagrams) \citep{Ado2015}.
By analogy, we  here evaluate the corresponding crossed-line contributions
for IPAHE in our  Dirac model with the warping term.

For each disorder class, the crossed impurity-line diagrams do not
generate new magnetic-field scalings or additional angular harmonics
beyond those in the noncrossing approximation. Specifically, for the
in-plane Hall conductivity independent of impurity density $n_{i}$,
the noncrossing contributions $\sigma_{xy,V_{i}}^{nc},i=\left(0,x,y,z\right)$
{[}i.e., Eqs. \ref{eq:V0} to \ref{eq: ipahe}{]} and crossed contributions
($\mathrm{X}$ and $\Psi$ diagrams) exhibit the same dependence on
the in-plane magnetization $\mathcal{M}_{\parallel}$ and the same
characteristic angular dependence:
\begin{align}
\sigma_{xy,V_{i}}^{nc},\sigma_{xy,V_{i}}^{X}+\sigma_{xy,V_{i}}^{\Psi} & \sim\begin{cases}
\mathcal{M}_{\parallel}^{3}\sin3\theta, & \left(i=0,z\right)\\
\mathcal{M}_{\parallel}\sin\theta, & \left(i=x,y\right)
\end{cases}.\label{eq: X Psi ipahe}
\end{align}

Quantitatively, the crossed and noncrossing contributions are of comparable
magnitude for the scalar ($V_{0}$), spin-conserving ($V_{z}$) and
spin-conserving ($V_{x}$ and $V_{y}$) disorder classes:
\begin{align}
\frac{\sigma_{xy,V_{i}}^{X}+\sigma_{xy,V_{i}}^{\Psi}}{\sigma_{xy,V_{i}}^{nc}} & \approx\begin{cases}
-0.8, & \left(i=0\right)\\
0.8, & \left(i=x,y,z\right)
\end{cases}.
\end{align}

Thus, the crossed diagrams mainly  make a substantial correction to
the amplitude, but not alter the angle dependence of the IPAHE. The
details of the formulation and numerical evaluation are provided in
the Supplemental Material (Sec. F) \citep{SMs}.

\section{In-plane magnetoconductivity}

\begin{figure}
\includegraphics[width=8.5cm]{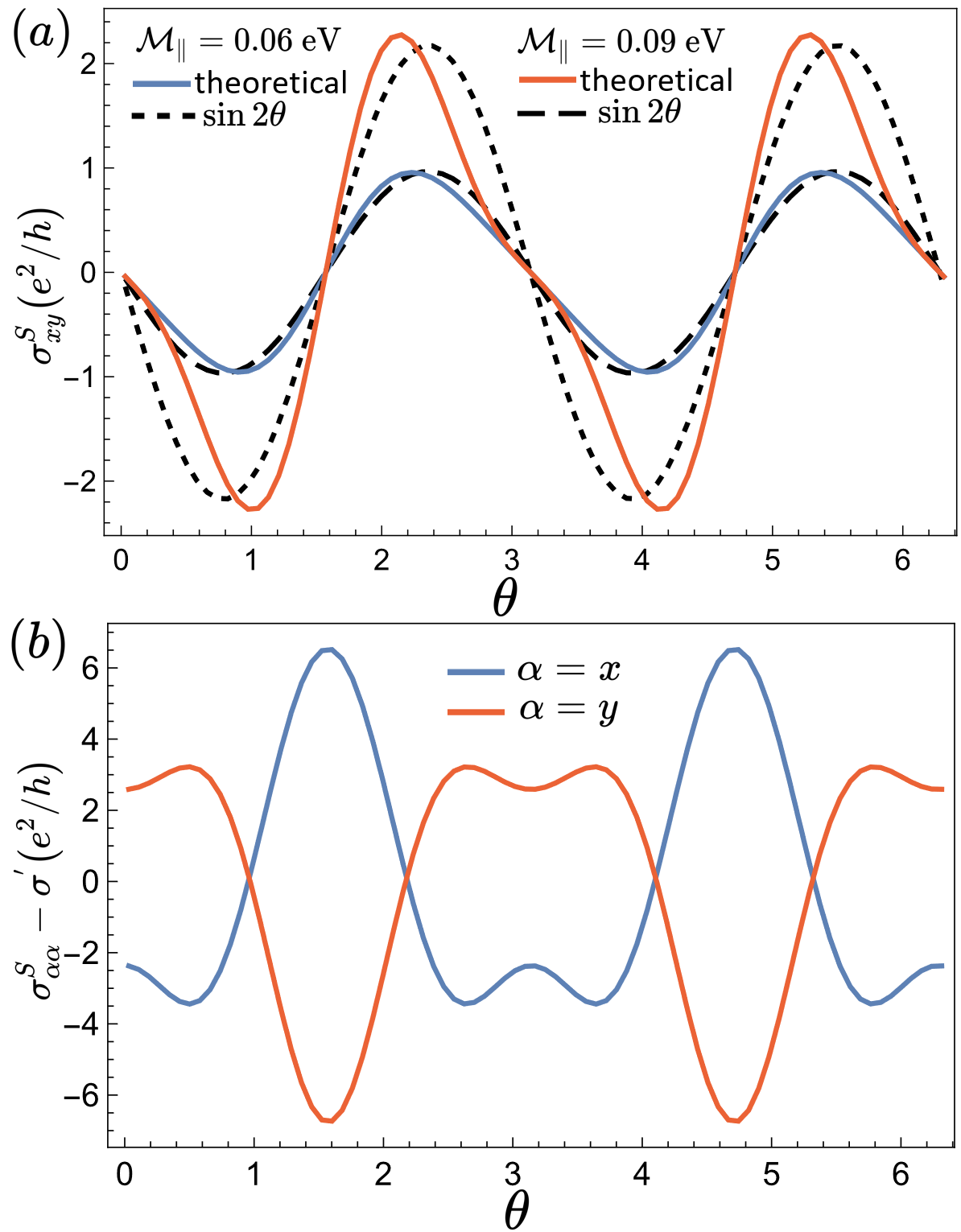}
\begin{turnpage}
\caption{\label{fig: nur} (a) Comparison between the calculated planar Hall
conductivity and the approximate expression ($\sin2\theta$ part in
Eq. $\left(\ref{eq:sigma S V0}\right)$) for two magnitudes of the
in-plane magnetization. (b) The longitudinal conductivities $\sigma_{\alpha\alpha}^{S}-\sigma^{'}$
with $\mathcal{M}_{\parallel}=0.12\ \mathrm{eV}$. Other parameters
are $\lambda=0.01\ \mathrm{eV\ nm^{3}},v=0.2\ \mathrm{eV\ nm},\ E_{F}=0.2\ eV,\ \tau=2\ \mathrm{ps}$.}
\end{turnpage}

\end{figure}
 To comprehensively understand some puzzling magnetotransport phenomena
in topological materials, we briefly discuss the impact of different
classes of impurity scattering on in-plane magnetoconductivity. It
is instructive to decompose the electric conductivity tensor into
the symmetric and antisymmetric parts $\sigma_{\alpha\beta}(\boldsymbol{\mathcal{M}})=\sigma_{\alpha\beta}^{S}(\boldsymbol{\mathcal{M}})+\sigma_{\alpha\beta}^{A}(\boldsymbol{\mathcal{M}})$,
where the antisymmetric part $\sigma_{\alpha\beta}^{A}=\frac{\sigma_{\alpha\beta}-\sigma_{\beta\alpha}}{2}$
reflects the dissipationless nature, and the symmetric part of the
off-diagonal components of electric conductivity $\sigma_{\alpha\beta}^{S}=\frac{\sigma_{\alpha\beta}+\sigma_{\beta\alpha}}{2}$
is usually related to the planar Hall effect (PHE) and has the essential
nature of anisotropic magnetoresistance. The corresponding Onsager
relation (Eq. $\eqref{eq:Onsager relation}$) becomes
\begin{align}
\sigma_{\alpha;\beta}^{A/S}(\boldsymbol{\mathcal{M}},\boldsymbol{u},\tau) & =-\sigma_{\alpha;\beta}^{A/S}(-\boldsymbol{\mathcal{M}},-\boldsymbol{u},-\tau).
\end{align}

We would like to further calculate the symmetric part of conductivity
by using the Kubo formula of the noncrossing diagrams in Fig.\ref{fig:second order sj-1}.
For the coexistence of the scalar impurities and magnetic impurities
(leading order in $\lambda$) (see Sec. E of the Supplemental Material
\citep{SMs}), 

\begin{align}
\sigma_{xy,V_{i}}^{S} & =\frac{e^{2}}{h}\frac{2\lambda\mathcal{M}_{\parallel}\cos\theta\mathcal{M}_{z}}{vn_{i}u_{0}^{2}}\gamma_{V_{i}},\label{eq: MzMin}
\end{align}
with $\gamma_{V_{0}}=3\left(E_{F}^{2}-\mathcal{M}_{z}^{2}\right)\left(E_{F}^{2}-9\mathcal{M}_{z}^{2}\right)/\left(E_{F}^{2}+3\mathcal{M}_{z}^{2}\right)^{2}$,
$\gamma_{V_{z}}=-3\left(E_{F}^{2}-\mathcal{M}_{z}^{2}\right)\left(7E_{F}^{2}+\mathcal{M}_{z}^{2}\right)/\left(3E_{F}^{2}+\mathcal{M}_{z}^{2}\right)^{2}$
and $\gamma_{V_{x}}=\gamma_{V_{y}}=-5$, where $\gamma_{V_{i}}$ is
the dimensionless coefficient produced by velocity correction under
different impurity classes, which is only related to the out-of-plane
magnetic field. The sign of this magnetoconductivity solely depends
on the direction of the in-plane magnetic field, and it has mirror-$y$
symmetry.

Motivated by the recent planar Hall experiments, we would study the
in-plane magnetoconductivity in the presence of the only in-plane
magnetic field in the Kubo formula approach. One keeps all terms in
the lowest order of the hexagonal warping $\lambda$ and has $\sigma_{xy}^{S}$
\begin{align}
\sigma_{xy,V_{0}}^{S}=\sigma_{xy,V_{z}}^{S} & =-\frac{e^{2}}{h}\frac{\lambda^{2}\mathcal{M}_{\parallel}^{2}}{v^{4}n_{i}u_{0}^{2}}\left(\frac{9}{2}E_{F}^{2}\sin2\theta\right.\label{eq:sigma S V0}\\
 & \left.+\frac{9}{4}\mathcal{M}_{\parallel}^{2}\sin2\theta+\frac{27}{4}\mathcal{M}_{\parallel}^{2}\sin4\theta\right),\nonumber \\
\sigma_{xy,V_{x}}^{S}=\sigma_{xy,V_{y}}^{S} & =-\frac{e^{2}}{h}\frac{\lambda^{2}\mathcal{M}_{\parallel}^{2}}{v^{4}n_{i}u_{0}^{2}}\left(\frac{9}{2}E_{F}^{2}\sin2\theta\right.\nonumber \\
 & \left.+\frac{15}{4}\mathcal{M}_{\parallel}^{2}\sin2\theta+\frac{33}{4}\mathcal{M}_{\parallel}^{2}\sin4\theta\right).\label{SigxyVxy}
\end{align}
One can see that, in Fig. \ref{fig: nur}(a), our analytical theory
could offer an alternative explanation of some unexplained features
of the longitudinal magnetoresistance {[}noticeable deviation from
the conventional part oscillating as $\sin2\theta$ (dashed lines){]}
in the Sn-doped topological insulator $\mathrm{Bi}_{1.1}\mathrm{Sb_{0.9}}\mathrm{Te}_{2}\mathrm{S}$
\citep{WuB2018apl,WangCM2023PRB}. Similarly, we retain the parts
that are quadratic in $\lambda$ and have the longitudinal conductivities
$\sigma_{\alpha\alpha}^{S}$ as
\begin{align}
\sigma_{xx,V_{0}}^{S} & =\sigma^{'}+\frac{e^{2}}{h}\frac{\lambda^{2}\mathcal{M}_{\parallel}^{2}}{v^{4}n_{i}u_{0}^{2}}\frac{1}{4}\left(\frac{4\mathcal{M}_{\parallel}^{4}\cos6\theta}{E_{F}^{2}}\right.\nonumber \\
 & \left.-9\left(\mathcal{M}_{\parallel}^{2}+2E_{F}^{2}\right)\cos2\theta+27\mathcal{M}_{\parallel}^{2}\cos4\theta\right),\label{SigxxV0}\\
\sigma_{yy,V_{0}}^{S} & =\sigma^{'}+\frac{e^{2}}{h}\frac{\lambda^{2}\mathcal{M}_{\parallel}^{2}}{v^{4}n_{i}u_{0}^{2}}\frac{1}{4}\left(\frac{4\mathcal{M}_{\parallel}^{4}\cos6\theta}{E_{F}^{2}}\right.\nonumber \\
 & \left.+9\left(\mathcal{M}_{\parallel}^{2}+2E_{F}^{2}\right)\cos2\theta-27\mathcal{M}_{\parallel}^{2}\cos4\theta\right),\label{SigyyV0}
\end{align}
where $\sigma^{'}=$$\left[e^{2}\lambda^{2}\left(-2\mathcal{M}_{\parallel}^{6}+5E_{F}^{6}+27\mathcal{M}_{\parallel}^{2}E_{F}^{4}+9E_{F}^{2}\mathcal{M}_{\parallel}^{4}\right)\right.$
$\left.\left.+E_{F}^{2}2v^{6}\right]\right/\left(2hE_{F}^{2}v^{4}n_{i}u_{0}^{2}\right)$
is independent of the magnetic field direction. We only consider the
scattering process without velocity correction, the corresponding
relaxation time is $\hbar/(2\tau)\simeq n_{i}u_{0}^{2}E_{F}/(4v^{2})$.
It should be emphasized that the analytical $\sigma_{xx}^{S}$ and
$\sigma_{yy}^{S}$ above as well as the result in Fig. \ref{fig: nur}(b)
could capture the key feature of the numerical result of anisotropic
magnetoconductivity of the Dirac surface states with hexagonal warping
term in \citep{Akzyanov2018PRB}, a superposition of contributions
with $\pi$ period and $\pi/2$ period. That is, the hexagonal warping
term offers a new mechanism for the fourfold symmetric anisotropic
in-plane magnetoresistance that is distinct from that due to the topological
orbital magnetic moment of Dirac fermions \citep{tu25}. Thus, our
theory enables us to well understand the relevant magnetotransport
experiments.  

\section{conclusion and discussions}

In summary, we mainly explored the extrinsic part of IPAHE based on
the 2D massive Dirac fermions with warping term. The distinct behaviors
of IPAHE against the three universal classes of disorder scattering
we found to be consistent with previous results with a noncross approximation
of massive Dirac fermions in two limits, and we found the crossed
diagrams did not qualitatively modify the angular dependence of the
IPAHE. Notably, the spin-flipping scattering could induce extrinsic
contributions of sinusoidal oscillations with periods of $\pi$ and
$2\pi$, in contrast to the standard 2D massive Dirac fermions. In
addition, we briefly calculated the in-plane magnetoresistance and
made some comparisons with previous results. Our work could provide
a comprehensive picture of IPAHE under general spin-dependent scattering
and help us understand the Hall transport of the quantum materials.
Moreover, our theory could be extended to the nonlinear Hall effect
\citep{PanHe2023prl,DimitrieCulcer2024prb,DimitrieCulcer2023prb}.
\begin{acknowledgments}
The authors thank Yang Gao and C. M. Wang for insightful discussions.
This work was financially supported by the National Key R\&D Program
of the MOST of China (Grant No. 2024YFA1611300), the National Natural
Science Foundation of China under Grants No. 12174394 and No. U2032164.
J.Z. was also supported by HFIPS Director’s Fund (Grants No. YZJJQY202304
and No. BJPY2023B05), Anhui Provincial Major S\&T Project (No. s202305a12020005)
and the Basic Research Program of the Chinese Academy of Sciences
Based on Major Scientific Infrastructures (grant No. JZHKYPT-2021-08)
and the High Magnetic Field Laboratory of Anhui Province under Contract
No. AHHM-FX-2020-02.
\end{acknowledgments}

\bibliographystyle{apsrev4-1}
\bibliography{IPAHEDisorder}

\end{document}